\documentstyle[12pt]{article}
\input{tcilatex}

\begin{document}

\title{Collective nature of $0^{+}$- states in deformed \\
rare-earth nuclei.}
\author{Vladimir P. Garistov \\
%EndAName
{\small Institute for Nuclear Research and Nuclear Energy,}\\
{\small Sofia, Bulgaria.}\\
}
\maketitle

\begin{abstract}
%======================= ABSTRACT ===================================
The description of the energiy spectra of 0$^{+}$ states for rare-earth
nuclei has been done involving the degree of the collectivity of
corresponding \ 0$^{+}$-state as a systematics parameter.
Holshtein-Primakoff representation leads to very good agreement with
experiment. Within the framework of this approach the parameter of the
collectivity is mainly determined by pairs of particles constructed on
single ''effective'' level. The results may be helpful both for nuclear
structure experimentalists and theorists in their investigations of
low-lying states structure and transition probabilities.
\end{abstract}

The great amount of experimental data for energy spectra and transition
probabilities \ evokes the necessity of simplified description that can be
easily used by experimentalists in explaining the collective properties of
states and their systematics. For example the peculiarities of the 0$^{+}$
spectra in nuclei and E0 transitions provoke great interest. In particular,
the experimental data of the E0 transitions from second 0$^{+}$- state to
the ground state in even-even Sn and Cd isotopes indicate that they are very
weak. The transitions from second 0$^{+}$ to the first 0$^{+}$ state in Cd
are weak too, but in Sn isotopes the corresponding transition probability is
very strong . Also the transitions from first to the ground 0$^{+}$ states
in Sn are weak, but in Cd are exceptionally strong \cite{1}.There are many
investigations dedicated to the $E0$ and $E2$ transition probabilities and
analyzing the 0$^{+}$ spectra in different nuclei \cite{2}. For instance in
the rare - earth region the values of the $\
B(E2;2_{K=0_{2}^{+}}^{+}->0_{g.s.}^{+})\;$and$%
\;B(E2;2_{K=0_{2}^{+}}^{+}->4_{g.s.}^{+})$ \ transitions \ as a function of
neutron number change drastically for different isotopes \cite{3}.In this
paper we study the low-energy \ 0$^{+\;}\;$spectra within the framework of
simplified pairing vibrational model in Holstein-Primakoff representation\
and investigate the systematics of the 0$^{+}\;$states for large amount of
rare-earth nuclei. We analyze the behavior of their energies using the
notion $k\;$- degree of the collectivity as a systematics parameter\ for the
corresponding \ 0$^{+}$- states.

By minimizing the Chi-square values of the energies for different
permutation of the possible values of $k\;$we obtain the spectra as a
function of this parameter $k$ \ for a large amount of 0$^{+}$states of
deformed rare earth nuclei. These energies can be explained by simple
formula : 
\begin{equation}
E_{n}=Ak-Bk^{2}  \label{å1}
\end{equation}
\ 

As the description of the structure of the 0$^{+}$ nuclear states in terms
of pair configurations continues to be of great interest both for theorists
and experimentalists we define the parameter of the degree of collectivity $%
k\;$as related to the monopole phonon operators 
\begin{equation}
\begin{array}{ll}
R_{+}^{j}= & {\frac{1}{2}}\sum\limits_{m}(-1)^{j-m}\alpha _{jm}^{\dagger
}\alpha _{j-m}^{\dagger }\;\text{; \ \ } 
\begin{array}{ll}
R_{-}^{j}= & {\frac{1}{2}}\sum\limits_{m}(-1)^{j-m}\alpha _{j-m}\alpha
_{jm}\;,
\end{array}
\\ 
& 
\begin{array}{ll}
R_{0}^{j}= & {\frac{1}{4}}\sum\limits_{m}(\alpha _{jm}^{\dagger }\alpha
_{jm}-\alpha _{j-m}\alpha _{j-m}^{\dagger })\;,
\end{array}
\end{array}
\end{equation}

where $\alpha _{jm}^{\dagger }$,$\;\alpha _{jm}$ are the nucleons creation
and annihilation operators.

Let us consider the Hamiltonian for $N$ particles placed on ''effective''
single level j $\;$in terms of the operators $R_{+}^{j}\;${\bf ,}$%
\;R_{-}^{j}\;${\bf , }$R_{0}^{j}$ :

\begin{eqnarray}
H &=&{\bf \alpha }R_{+}^{j}R_{-}^{j}+{\bf \beta }R_{0}^{j}R_{0}^{j}+\frac{%
{\bf \beta }\Omega }{2}R_{0}^{j}  \label{h1} \\
\Omega &=&\frac{2j+1}{2}  \nonumber
\end{eqnarray}

The operators (\ref{RR}) satisfy the commutation relations :

\begin{equation}
\left[ R_{0}^{j},R_{\pm }^{j}\right] =\pm R_{\pm }^{j}\text{; \ \ \ \ \ \ \ }%
\left[ R_{+}^{j},R_{-}^{j}\right] =2R_{0}^{j}  \label{cr1}
\end{equation}

In order to simplify the notations further we will omit the indices$\;j$.

Let us now present this Hamiltonian in terms of \ ''ideal'' boson creation
and annihilation operators $\ \;\;\left[ b,b^{\dagger }\right] =1\ ;\;\;\;%
\left[ b,b\right] =\left[ b^{\dagger },b^{\dagger }\right] =0\ $\ , using
the Holstein-Primakoff transformation \cite{5} for the operators $R_{+\text{%
, }}R_{-}\;$and $R_{0}$ : 
\begin{equation}
\begin{array}{ccc}
R_{-}=\sqrt{\Omega -b^{\dagger }b}\;b\text{; \ \ \ } & R_{+}=b^{\dagger }%
\sqrt{\Omega -b^{\dagger }b}\text{; \ \ \ } & R_{0}=b^{+}b-\Omega /2
\end{array}
\label{HP}
\end{equation}

The transformations ( \ref{HP}\ ) conserve the commutation relations (\ \ref
{cr1}) between $R_{+}$, $R_{-}$ and $R_{0}\;\;$operators \ . Thus for the
Hamiltonian (\ref{h1}) in terms of the new boson creation and annihilation
operators $b^{+},\;b$\ we have: 
\begin{equation}
H=Ab^{\dagger }b-Bb^{\dagger }bb^{\dagger }b  \label{Hbos}
\end{equation}
\ $\ $where :

\begin{equation}
\begin{tabular}{l}
$A=\alpha (\Omega +1)-\beta \frac{\Omega }{2}$ \\ 
$B=$\ $\alpha -\beta $%
\end{tabular}
\label{param}
\end{equation}

The energy of any monopole excited state\ \ $\left| n\right\rangle =\frac{1}{%
\sqrt{n!}}(b^{+})^{n}\left| 0\right\rangle ;\;$where $b\left| 0\right\rangle
=0\;\;$can be written as: \ 
\begin{equation}
E_{n}=An-Bn^{2}  \label{en}
\end{equation}

$\bigskip \ $We see that we have the same energy spectrum as spectrum ( \ref
{å1} ) ($k\rightarrow $ $n$)

The parameters of the approach which we have used are presented in figure 1
along with the experimental and calculated 0$^{+}$state distributions . The
calculated energies \ are distributed \ in bell form because of the
anharmonic terms in the Hamiltonian (\ref{Hbos})\ and often the lowest 0$%
^{+}\;$states have much more collective structure ( bigger $k$ ) than the
states with higher energies. In the framework of this simple model we can
predict that additional 0$^{+}$states should exist in the following cases: $%
\;$\ $^{194}Pt\;$(one phonon state - 0.75 MeV, six phonon state - 2.38 MeV,
and seven phonon state -\ 2.1 MeV), $^{196}Pt\;$(one phonon state - 0.6 MeV,
three phonon state - 1.5 MeV, twelve phonon state -\ 1.0 MeV ), $^{188}$Os
(one phonon state - 0.75 MeV and two phonon state - 1.3 MeV ) and $^{158}$Er
(one phonon state - 1.2 MeV). We indicate these predicted states by ''?'' in
the figures 1$^{a}$ -\TEXTsymbol{>} 1$^{d}$. Thus it may be interesting to
measure $E0\ $\ transition probabilities in these nuclei and especially in \
\ $^{194}Pt\;$nucleus from one phonon 0$^{+}$ state with energy 0.75 MeV to
the ground state, in \ \ $^{196}Pt\;$nucleus from one phonon 0$^{+}$ state
with energy 0.6 MeV to the ground state,\ in $^{188}$Os from one phonon 0$%
^{+}\;$state - 0.75 MeV to the ground state and in $^{158}$Er from one
phonon 0$^{+}$ state - 1.2 MeV to the ground state. We expect relatively
intensive $E0\ $\ transitions for these cases. Experimental data about the
rotational bands in deformed nuclei show that the dependence of the energy
on angular momentum$\;L$ is qualitatively similar for the ground band and
the bands constructed on any excited $0^{+}$\ state . So in the first
approximation one may consider the rotational bands constructed on different
excited $0^{+}\;$states without including the band head structure.
Nevertheless the influence of $0^{+}$ states structure on the rotational
spectra must be included in order to explain the small quantitative
differences in rotational bands with different $0^{+}$ band heads as well as
transition probabilities,for instance the peculiarities in $%
B(E2;2_{K=0_{n}^{+}}^{+}->0_{g.s.}^{+})\;$\cite{3}. This investigation is in
successful progress.$\ $Furthermore the results of this paper may be helpful
for more sophisticated analysis of the collective structure of the low-lying
nuclear states. Having in mind the results of this paper one can estimate
directly the degree of collectivity of any 0$^{+}$ excited state. $\ \ \ $

$\ \ $ I would like to thank professors S. Pittel, M.Stoitsov,
Ani.Aprahamian, Ana Georgieva and P. Terziev for fruitful discussions and
help.

This work has been supported in part by the Bulgarian National Foundation
for Scientific Research under project $\Phi $ - 809.\qquad\

Caption to the figure:\\

Figure 1. {\small Comparison of calculated with Hamiltonian ( \ref{h1}, \ref
{Hbos}) and experimental \cite{5} }$0^{+}${\small \ state energies for rare
- earth nuclei }$^{156}${\small Gd , }$^{188}${\small Os, }$^{194}${\small 
Pt, }$^{196}${\small Pt.}

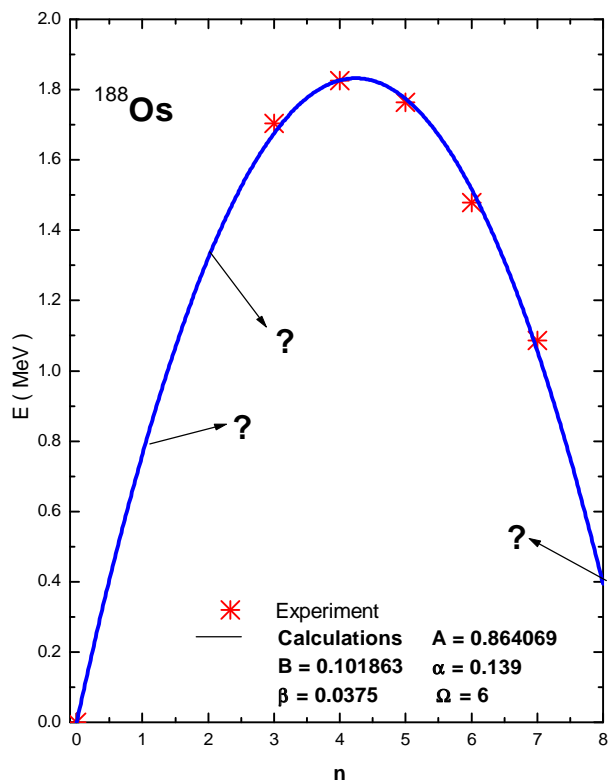
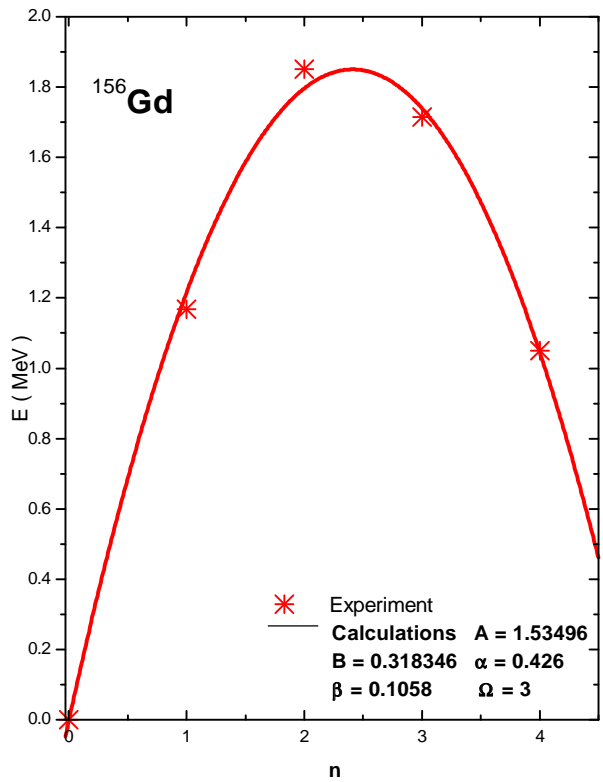
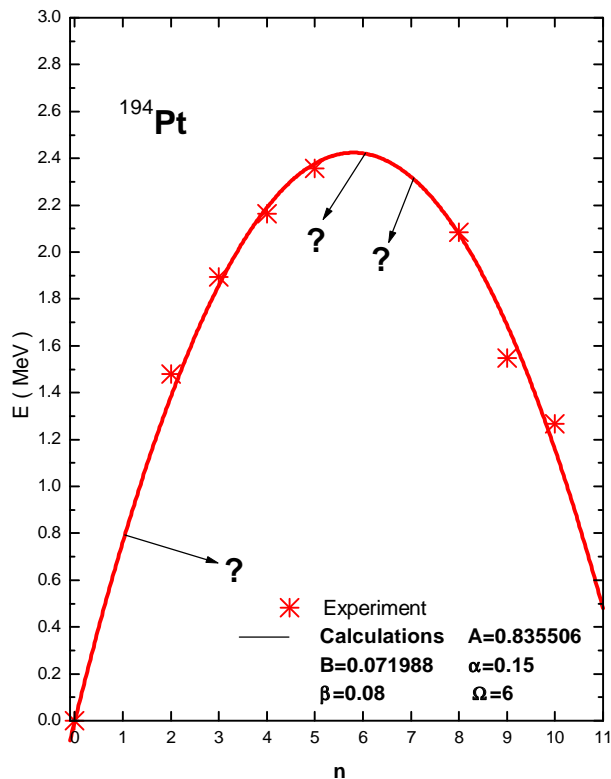
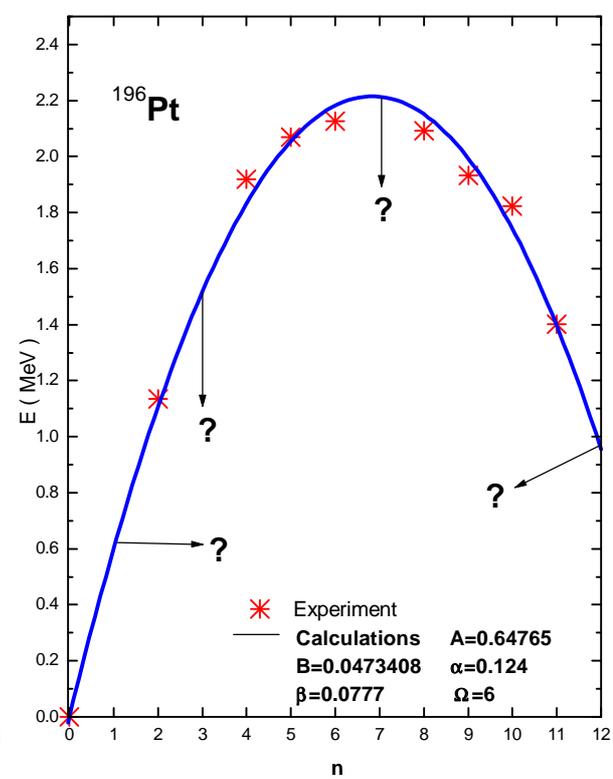
\end{document}